\documentclass[conference]{IEEEtran}
\ifCLASSINFOpdf
   \usepackage[pdftex]{graphicx}
   \DeclareGraphicsExtensions{.pdf,.jpeg,.png}
\else
\fi

\usepackage{algorithm}

\usepackage{algpseudocode}
\usepackage{cite}
\usepackage[final]{changes}

\definechangesauthor[name={Dominik}, color=red]{dominik}
\definechangesauthor[name={Suat}, color=blue]{suat}
\definechangesauthor[name={Mumin}, color=purple]{mumin}
\definechangesauthor[name={Enes}, color=purple]{enes}

\usepackage{atbegshi,picture}
\usepackage{lipsum}

\AtBeginShipout{\AtBeginShipoutUpperLeft{%
  \put(\dimexpr\paperwidth-1cm\relax,-1.5cm){\makebox[0pt][r]{\framebox{This paper will be presented in IEEE ICC BIoTCPS, 2020, Dublin, Ireland}}}%
}}

\begin{document}
%
\title{Improving Sustainability of Cryptocurrency Payment Networks for IoT Applications}

\author{\IEEEauthorblockN{Suat Mercan, Enes Erdin, and Kemal Akkaya} 
\IEEEauthorblockA{Dept. of Elec. and Comp. Engineering, Florida International University, Miami, FL 33174\\ Email: \{smercan, eerdi001, kakkaya\}@fiu.edu}
}

\maketitle

\begin{abstract}

Blockchain-based cryptocurrencies received a lot of attention recently for their applications in many domains. IoT domain is one of such applications, which can utilize cryptocurrencies for micorpayments without compromising their payment privacy. 
However, long confirmation times of transactions and relatively high fees hinder the adoption of cryptoccurency based micro-payments. The payment channel networks is one of the proposed solutions to address these issue where nodes establish payment channels among themselves without writing on blockchain. IoT devices can benefit from such payment networks as long as they are capable of sustaining their overhead. Payment channel networks pose unique characteristics as far as the routing problem is concerned. Specifically, they should stay balanced to have a sustainable network for maintaining payments for longer times, which is crucial for IoT devices once they are deployed. In this paper, we present a payment channel network design that aims to keep the channels balanced by using a common weight policy across the network. We additionally propose using multi-point connections to nodes for each IoT device for unbalanced payment scenarios. The experiment results show that we can keep the channels in the network more equally balanced compared to the minimal fee approach. In addition, multiple connections from IoT devices to nodes increase the success ratio significantly.

\end{abstract}
\begin{IEEEkeywords}
Bitcoin, Blockchain, Payment channel network, Internet of Things, Routing, Lightning Network
\end{IEEEkeywords}

%
\maketitle

\section{Introduction}

Internet of Things (IoT) have been adapted in many domains, but its great potential has not been unleashed yet because of various issues such as scalability, security, privacy, connectivity, etc. 
Another venue where IoT devices are heavily utilized is the commerce business. When a good or service is sold (e.g., vehicle charging, parking payment, vending machine purchase, etc.), the IoT devices will eventually need to send and receive payments.
Cryptocurrency based payments \cite{bitcoin,ethereum} provide a higher level of privacy and security for both parties as it hides the payee and payer identities, prevent frauds since the transactions are secured using cryptographic techniques, and Blockchain ledger provides non-repudiation in case of conflicts.

In Blockchain technology \cite{blockchainapp} a consensus mechanism (e.g. Proof of Work, Proof of Stake, etc.) is employed which eliminates the necessity of a central authority to approve and keep the records. Since the Blockchain ledger is irreversible and anyone can have a copy of it, the failure of a participant does not cause the system to collapse. In this way, it not only removes the problem of single point of failure but also enables secure transactions in a trustless environment.
The distributed structure of the Blockchain is the source of the strength it holds, which unfortunately becomes the point of weakness in scalability \cite{gilad,croman} when it comes to increased number of users and payments. 

Specifically, the design of Bitcoin makes it inherently time-consuming and slow. For instance, by design, it takes around 10 minutes to add a new block to the Bitcoin.
Moreover, the limit in the block size also affects the performance. Not only these design parameters but also the hardware and bandwidth restrictions limit the number of transactions that can be completed in a time frame. The theoretical maximum number of transactions is calculated as 7 per second~\cite{tranrate} which is far lower than what Visa or MasterCard can process\cite{visatran}.
Additionally, the transaction fee, which can surge on congested days \cite{bitcoinfee}, is disproportionate to the amount to be sent.
Consequently, high transaction fees and long block confirmation times are two main issues preventing virtual currencies from scaling and being adopted for micro-payments that can be used in daily life.

\begin{figure}
    \centering
    \includegraphics[width=0.7\linewidth]{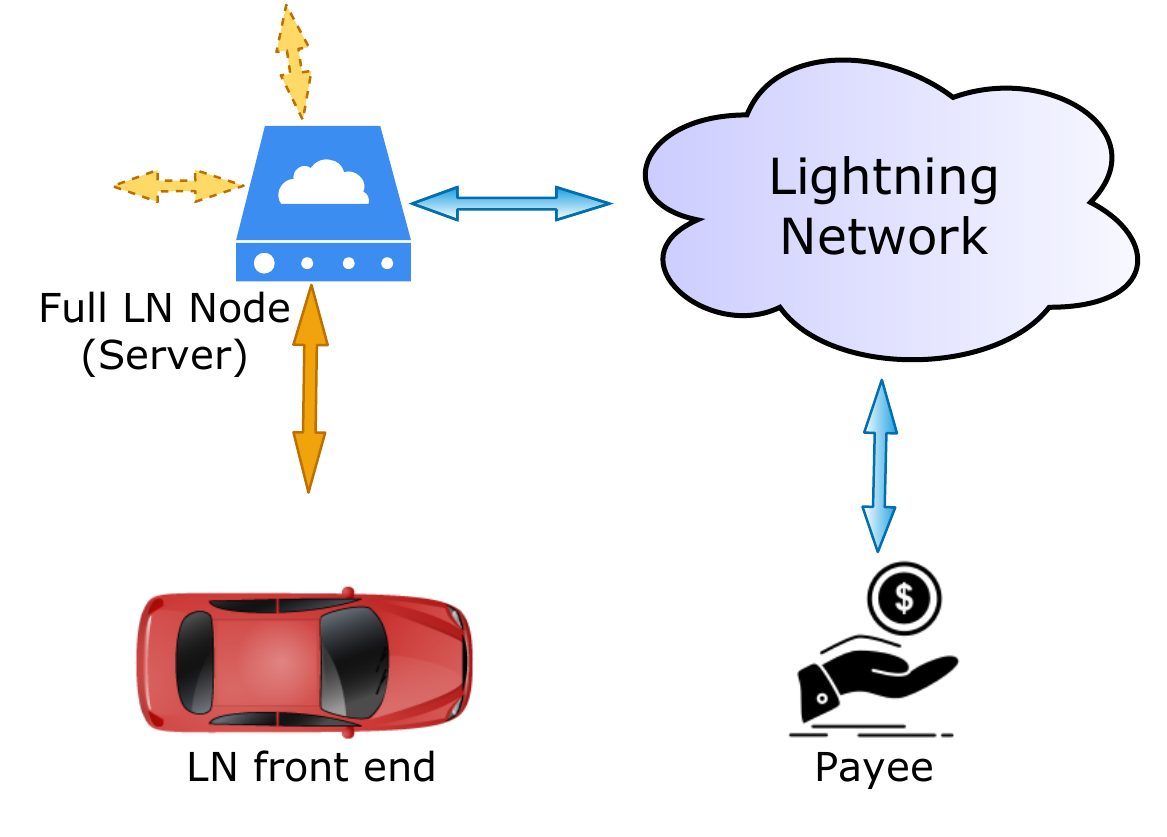}
    \vspace{-3mm}
    \caption{A conceptual design showing an IoT device making a payment in LN.}
    \label{fig:IoTLN}
    \vspace{-6mm}
\end{figure}

A payment channel network (also known as off-chain networks) \cite{ln,raiden,ripple} is one of the solutions proposed to address these problems of virtual currencies. It leverages the smart-contract concept to avoid writing every transaction on the blockchain. Instead, the transactions are done \textit{off-chain}. Basically, once a channel is created between two parties, an infinite number of transactions can be performed in both directions as long as there is available funds. Furthermore, since opening a channel is a costly operation in terms of time and money, nodes in a payment channel network are enabled to send payments to any node through other existing nodes by paying a \textit{transaction fee}. This forms an overlay network to represent the payment channel network where the balances among the nodes are considered as links. Recent Lightning Network (LN) is a perfect example of this concept that reached to almost 10K users in 2 years \cite{ln}. This payment network has nodes which charge transaction fees to users passing their data over them. 
Applying the PCN concept in the IoT domain will bring new opportunities for the users and the businesses. In the conceptual design shown in Fig. \ref{fig:IoTLN}, as an example, a car (light node) can make a payment via connecting to the LN through a full node. This design which constitutes a base for the startup company Breez \cite{breez} can be extended to any use case as the full LN node 
manages the LN protocol.

While payment channel network establishment is a promising method to solve the scalability challenge of blockchain-based cryptocurrencies for micro-payments, it has its own challenges related to operational efficiency, management and routing, etc.
One of the main characteristics of these payment channel networks is the way their channel capacities are consumed which raises new challenges in terms of insufficient balances in channels during routing. For instance, LN\cite{ln} is the most prominent implementation of this concept where the nodes are free to set the transaction fees which are the basis along with channel capacity to determine the routes from senders to receivers. Obviously, the users select the routes that have available capacity with minimal fees. Such selection causes exhaustion of available funds in one direction which might lead to a poorly connected or even a disconnected (i.e., partitioned) graph. Consequently, unidirectional payments and ignorance of imbalanced channels result in non-conductive nodes which impacts the overall efficiency of payment routing. As an example, in one of the recent studies, the researchers found that chance of sending a \$5 payment is around 50\% in LN \cite{diar} which is not acceptable for users.  Unfortunately, this problem has not been explored well as the attention is rather given mostly to routing mechanisms.



We argue that a \textit{balance-aware} routing can improve the stability and success ratio of the overall network in the long run. In this paper, we propose the adoption of two specific techniques to address the issues related to imbalance routing in payment channel networks. First, we use a common channel \textit{weight} policy to be adapted by all nodes instead of letting users individually decide. Weight here refers to a metric based on channel balance that is used for finding the least cost path using Dijkstra's algorithm. Fundamentally, the motivation is to encourage nodes to use high-balanced channels and avoid low-balanced ones for payments dynamically. This will help to utilize the channels in a manner that will keep available balances in the channels in all directions. Second, we suggest to exploit multiple ingress points to the network from customers (i.e., IoT devices) in order to further improve the solution. In this way, an IoT will have an option to choose the node to initiate a payment. Multiple entry point will be helpful especially in case of skewness in the payment flow. These two features will create symmetrically balanced channels and improve the efficiency of the payment channel networks. We implemented and tested the effectiveness of the proposed approach under various payments scenarios and observed that the payment routes can be significantly balanced. 


This paper is organized as follows: Section II summarizes the related work in the literature while Section III provides some background explaining the concepts used in payment networks and our assumptions.  Section IV presents the problem definition and our approach. In Section V, we assess the performance of the proposed mechanism. Finally, Section VI concludes the paper.

\section{Related Work}

Various methods of payment channel networks have been suggested and some are already being implemented. Lightning Network (LN)\cite{ln} for Bitcoin and Raiden\cite{raiden} for Ethereum are two examples in practice. LN is the most active one with more than 10000 nodes and 30000 channels \cite{acinq}. In LN source-routing is utilized for sending payments. A node first finds a path with available channel capacities, then initiates the transaction. Spider\cite{spider} applies packet-switching routing techniques to payment channel networks. The payments are split into micro-payments similar to MTU in computer networks. It uses congestion control and a best-effort model to improve payment throughput. It specifically chooses the paths that re-balances the channels. The payments are queued at spider routers and they are transferred when the fund is available. Flash\cite{flash} applies a distributed routing algorithm to better handle constantly changing balances. It differentiates mice and elephant payments. Small payments are sent randomly over pre-computed paths. For large payments, it probes the nodes to find the channel with available funds. Then it splits the payment into several smaller chunks. Revive\cite{revive} supposes that a node has multiple connections and the skewed payments make some of the links depleted. It tries to find cycles in the network and a user sends a payment to herself to re-balance the depleted channel through others. SpeedyMurmur\cite{speedymurmur} is focusing on the privacy of the payments by using an embedding-based routing algorithm. SilentWhisper\cite{silentwhisper} and Flare\cite{flare} are using landmark routing in which only some nodes store routing tables for the complete network. The rest only knows how to reach one of those landmark nodes. A user transmits the payment to the gateway node which handles the rest. Our work is focusing on a path selection strategy to choose the appropriate route for payments so that the overall network will be more sustainable (i.e., more transactions) as we leave the details of routing to other works.

\section{Preliminaries and Assumptions}

\textbf{Blockchain and Bitcoin:} Blockchain is a distributed database in which the building data structure is a block. For Bitcoin, a block is comprised of transactions (data), timestamp, nonce, the hash of the block and the hash of the previous block\cite{bitcoin}. The cooperation of honest nodes is the guarantee for the consistency of the blockchain. In cryptocurrency-based networks, the nodes come to a consensus for a transaction block by proving that they have enough interest in the network. Forinstance, in widely used Bitcoin Hashcash, proof of work (PoW) is utilized. The basic mechanism is that a miner node packs the transaction requests into a block and starts to calculate the hash of that block. The resulting hash should be smaller than a value which is again calculated based on the cumulative computational power of all of the miners. By changing the nonce value the miner aims to find a suitable result. Soon after a block is found, it gets distributed to the other nodes. When other nodes approve that block, the next block calculation starts.

\textbf{Off-Chain Payment Channels:} For Bitcoin, the average time of approval of a block is around 10 minutes. This time period puts a burden on the usability and practicality of the Bitcoin. Namely, the usage of Bitcoin for simple daily spending becomes almost impossible. The reason for that is a payee, as a heuristic, waits at least 6 blocks to count a transaction as a valid one. 


To solve that problem, thanks to the introduction of the smart contract mechanism to the blockchain, developers came up with the idea of the mechanism called ``off-chain payment channel''. In that mechanism, 2 users, say A and B, come to a mutual agreement on making a business. Then they sign a contract by transferring collateral to a shared 2-of-2 multi-signature address and initiate the channel by publishing it on the blockchain. This contract type is called ``Hash Time Locked Contracts'' (HTLC). When the users agree on any amount of payment they prepare a new HTLC, exchange the new contract, and update the state of the channel. To initiate a payment from a debtor a challenge, namely, a pre-image 864
is sent to the recipient. If the recipient can reply successfully to the challenge, the contract becomes valid, so, the ownership of the money is transferred. Off-chain mechanism brings a huge advantage that the peers do not need to publish every transaction on the blockchain. That is, the payments are theoretically instantaneous. Moreover, as there is no need for frequent on-chain transactions, the transactions will be protected from instantaneous unexpected high transaction fees. 

\begin{figure}[htb]
    \centering
    \includegraphics[width=0.7\linewidth]{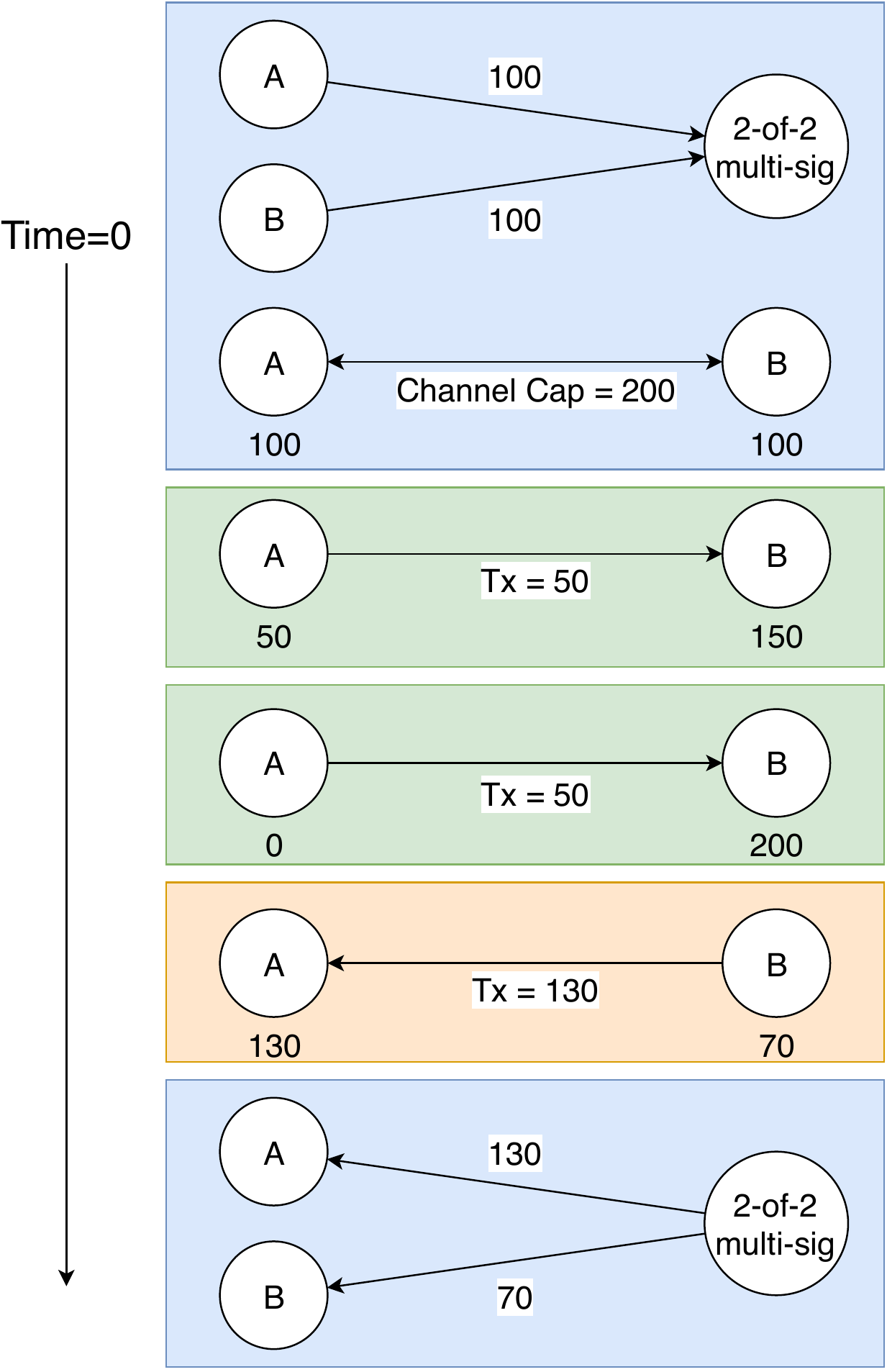}
    \caption{Illustration of a payment channel.}
    \label{fig:channels}
    \vspace{-5mm}
\end{figure}

An important feature of this type of channel is the direction of the payments matter. Specifically, two flows from opposite directions on the same link negate each other's capacity usage. This is shown in Fig. \ref{fig:channels}. At Time=0, a channel between two parties A and B is established. Both A and B put 100 unit of currency which in turn makes the channel capacity 200 units. After A makes 2 transactions each of which is 50 units, the directional capacity from A to B will be zero. Hence, A can not transfer any more unless B transfers his/her back some money. B sends 130 unit back and when they close the channel they get their corresponding shares from the multi-signature address.

\begin{figure}[htb]
    \centering
    \includegraphics[width=0.9\linewidth]{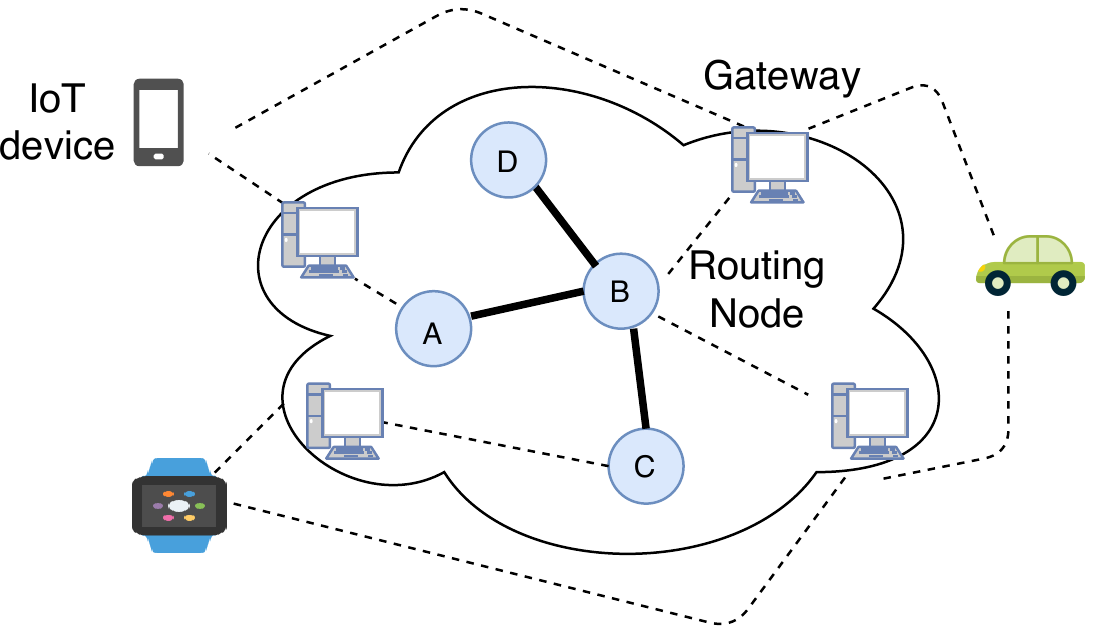}
    \caption{Payment channel network.}
    \label{fig:PBFT}
    \vspace{-3mm}
\end{figure}

\textbf{Payment Channel Networks:} Off-chain payment channels can be extended to a payment channel network idea. As shown in Fig.~\ref{fig:PBFT}, assume that A and B have a channel, and B and C have a channel too. If somehow, A wants to trade with C only, what s/he has to do is hash-lock a certain amount of money and forward it to C through B. As C already knows the answer to the challenge, C will get her/his money from B by disclosing the answer. The brilliance of the HTLC appears here. As C discloses the answer to the challenge, B learns the answer. Now, B will reply to the challenge successfully and get her/his share from A. In this way, one can reach everyone in a network through multi-hop payments forming a payment channel network. The customers can connect from any gateway to this payment network. In our case, IoT devices can pick one or more gateways to open an offchain channel through a wireless communication. 

\textbf{Assumptions}: In this paper, we assume that a payment channel network consists of nodes connected with off-chain payment channels. The channel capacity represents the amount of money deposited in a 2-of-2 multi-signature address. Although any node can send to and receive payment from any other node, we want to distinguish nodes as customers (IoT devices) and stores since we want to simulate the case that IoT devices want to use cryptocurrency for micropayment where the cash flow is mostly from a customer to a store for a service. 

We build up our work on a presumed existing routing protocol which is used by the nodes to advertise the weights to the rest of the network and carry a specified payment from a source to destination. We are not focusing on the efficiency and overhead of the routing protocol as our goal is to determine the location to start the payment and finding the most appropriate route. Each node is assumed to have the complete topology to calculate the path. 


\section{Proposed Approach}


In this section, we explain the motivation behind our approaches and then detail our solutions. 

\subsection{Problem Motivation and Overview}

\begin{figure}[h]
    \centering
    \includegraphics[scale=0.8]{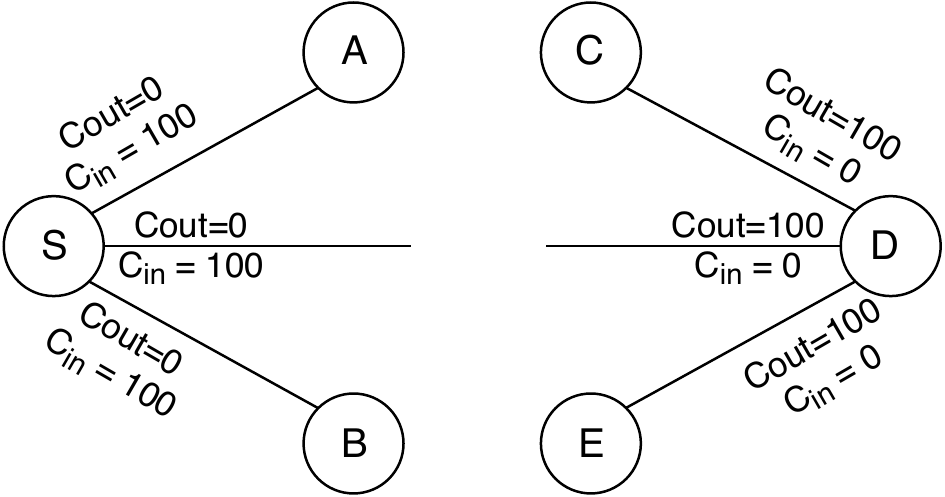}
    \caption{Depleted channels causes disconnection.}
    \label{fig:endnodes}
    \vspace{-3mm}
\end{figure}

A payment originates from a node, gets transmitted through intermediary nodes and arrives at the destination. The nodes are acting either as transit nodes or end-nodes. A node which is the source or destination for payment is called an end-node, and it is called a transit node if it only transfers from one neighbor to another. During this process,  various problems might occur causing the payment to fail or leading to inefficiency in transmission. This is not acceptable in many IoT applications where payments need to be done real-time and the service should be available at all times. We discuss these problems separately below:  
\vspace{0.2cm}


\noindent \textit{Problem: Highly Directional Payments:}  If a node constantly transmits payments in one direction on the same channel, the balance of the channel will be depleted in that direction. It will create a loose connection or disconnection in the network which might cause 1) a group of nodes be disconnected from the rest of the network until the balance of the channel is increased, 2) a node to have to travel longer paths, 3) two payments arriving a node simultaneously not get transmitted due to lack available fund. For instance, S can not send payment to A in Fig.~\ref{fig:endnodes}. In current payment networks, users are generally calculating routes based on the optimum fee which are set by node owners. They are not subject to any rule when they are setting the fees. The user optimal flow may be different than the system's optimal flow.

\vspace{0.1cm}
\noindent \textit{Problem: Over-used nodes:} The second problem is regarding the originator and receiver of a payment. If a store has used up its outgoing capacity, then this node can not initiate any payments and function as a transit node. Similarly, if a store consumed all of its incoming capacity, then it can not receive payment and function as a transit node as well.

\begin{figure}[!h]
    \centering
    \includegraphics[width=.80\linewidth]{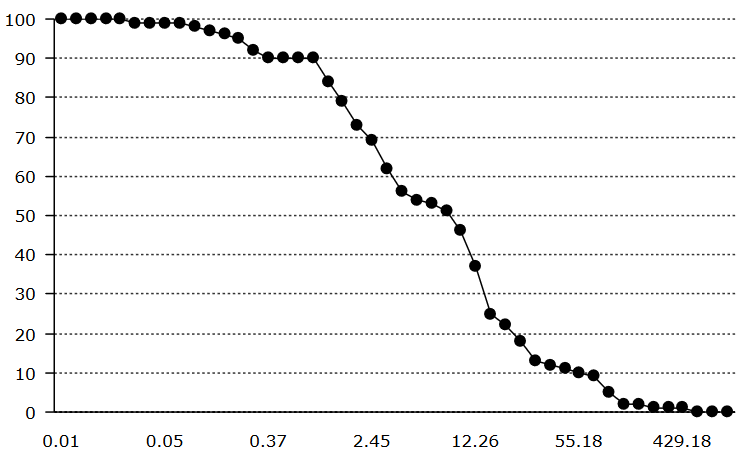}
    \caption{Success rate vs amount of the payment \cite{diar}.}
    \label{fig:diar}
    \vspace{-3mm}
\end{figure}

In Fig.~\ref{fig:endnodes}, IoT devices connected to S can not make payments through this node. They have to wait for S to receive a payment destined to it. A can not send payment to B through S. On the other hand, D can not receive any payment, it has to wait for any IoT device send payment originating from D. Similar to other cases, C can not send payment to E. We cannot control where the payment is destined but we can control the point that the payment might originate to some extent by using multiple connections from IoT devices to stores. A node receiving a high volume of payments should also be preferred as the source, and a node which has less outgoing capacity should be avoided to originate payments. Fig. \ref{fig:diar} shows the results of an experiment on the success rate of sending a transaction versus the amount being sent in USD in 2018 \cite{diar}. At the time of the experiment the mean capacity per channel in LN is reported to be \$20. The graph tells us that, with a success rate of 90\%, even sending \$1 is not guaranteed.

\vspace{-2mm}

\subsection{Proposed Solutions}
We now offer two solutions to address the aforementioned problems: 

\noindent \textit{Proposed Solution 1: Balance-aware Routing:} One of the most important factors for the success of the payment channel networks lies in keeping the channels balanced. We propose using a mandated weight calculation method which must be adopted by all the nodes in the network. Basically, each directed edge is assigned a weight inversely proportional to its current capacity. The route calculation will be based on this newly assigned weight. This will help keep the channels equally balanced in both ways and the overall network. Specifically, the weight of each channel is computed using the following equation by each node:
\vspace{-1mm}
\begin{equation}
    W = (MC - C_o)^2
\end{equation}

where $W$ represents the weight of a channel, $MC$ is the maximum allowed channel capacity set centrally, and $C_o$ is the current balance on the outgoing channel. This new weight adjusts the weight according to the current balance. The node (user) then calculates the shortest path using these newly computed weights to send the requested payment. Note that squaring the difference will significantly increase or decrease the weight. Consequently, this new weight will strongly encourage users to use channels with available balance while helping them to avoid routing over low-balanced channels.  Transmitting all the payments through high-balanced channels in this way will solve the balance inequality problem which is the key element for a successful payment network design. 
\vspace{0.1cm}


\noindent \textit{Proposed Solution 2: Utilization of multi-connection:} The IoT devices are to be connected to more than one node in the payment network so that they can initiate the payment from various vantage points. The route to the destination is calculated from all these points and the one that will contribute to network stability the most is picked.

\vspace{-2mm}
\begin{figure}[h]
    \centering
    \includegraphics[scale=0.75]{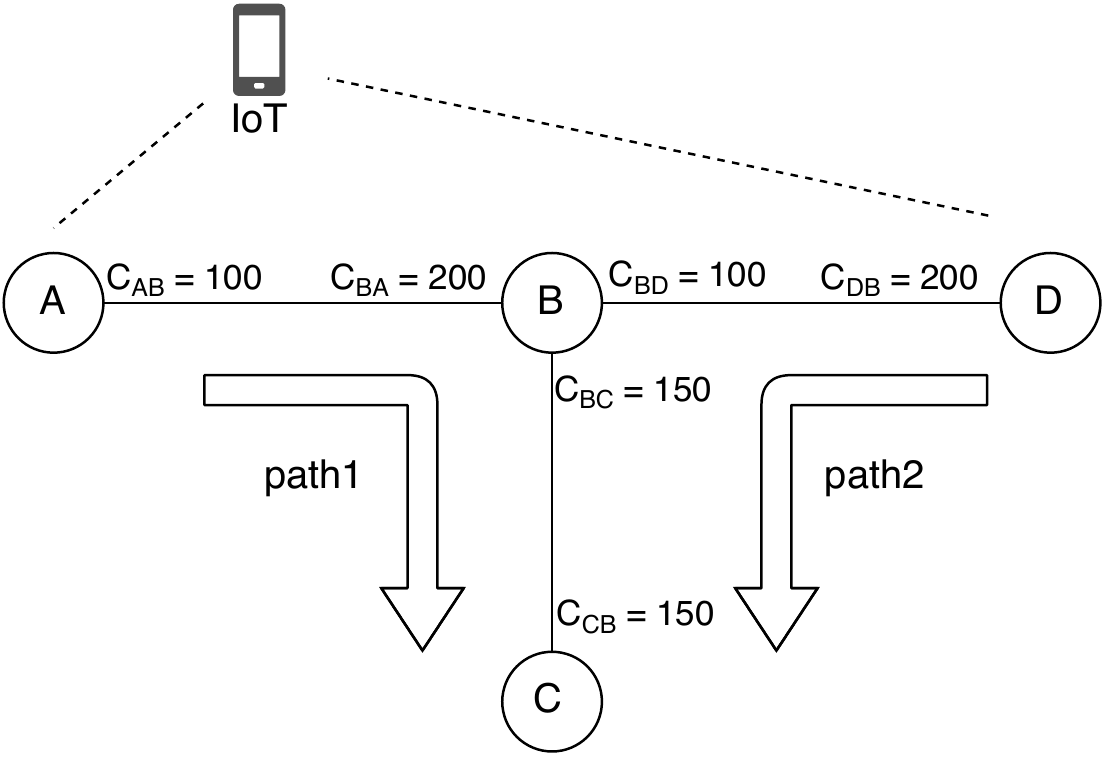}
    \caption{Multi-point connection enables re-balancing.}
    \label{fig:multipoint}
    \vspace{-2mm}
\end{figure}

For instance, customer (IoT device) shown in Fig.~\ref{fig:multipoint} has two options to initiate the payment to C: 1) $path1$ which starts from $A$ and 2) $path2$ which starts from D. Per our approach, the IoT device will choose $path2$ after the route calculation. This choice will help node $A$ to preserve its limited outgoing capacity ($C_{AB}$) and keep the channel between $A$ and $B$ balanced in both directions. It will also increase $D$'s incoming capacity, $C{BD}$ (due to payment sent to $B$) and more equally balance the channel between $B$ and $D$.


\begin{algorithm}
\caption{Route Calculation}\label{alg:routecalc}
\begin{algorithmic}[1]
\State Input: \textit{C=Store connection list}, \textit{G=Connected directed graph}
\For{every edge, $e$, in $G$}
    \State $G_e$.$weight$=($MC$-$G_e$.$balance)^2$
\EndFor
\hspace{2mm}// \textit{weight calculations are done}
\State min = Integer.Max
\For {every \textit{connection}, $s$ in $C$}
// \textit{Calculate shortest path from each point}
    \State $Path$=$ShortestPath$($G$, from=$s$, to=$d$) 
    \If{$Path$ less than $min$}
        \State $min$ = $Path$
    \EndIf
\EndFor
\State Output: $min$
\end{algorithmic}
\end{algorithm}

Overall, the proposed route calculation is given in Algorithm \ref{alg:routecalc}. The weight for each link for the outgoing connections is calculated first using Equation 1. Then based on the link weight, each node computes the shortest path to destination from its available connections using Dijkstra's shortest path algorithm. From amongst these, the minimum cost path is selected to initiate the payments. 

An IoT device basically selects one of these nodes as a connection point to make its payments. Our algorithm picks these connection points randomly based on the available number of connections allowed.

\section{Performance Evaluation}
We have performed extensive experiments to understand the impact of the proposed method on the performance as detailed in this section.  

\subsection{Experiment Setup}
We developed a simulator in Java that allows us to run the experiments and measure the defined metrics. There are various parameters to be set when running the experiments as listed in Table \ref{table1}. We also explain the details of dealing with other configurations. 
\vspace{-1mm}
\begin{table}[h]
\renewcommand{\arraystretch}{1.5}
\caption{Experiment Parameters}
\label{table1}
\centering 
\begin{tabular}{|c|c|}
\hline
 Number of Nodes           & 100     \\
\hline \hline
 Node degree               & 3       \\
\hline \hline
 Initial Channel Capacity  & 50-150  \\
\hline \hline  
 Payment amount            & 5-15    \\
\hline \hline
 Number of Payment         & 5K      \\
\hline
\end{tabular}
\end{table}
\vspace{-1mm}
\textit{Network configuration:} In the first set of experiments, we assume that each IoT device is connected to only one store. This allows us to evaluate the impact of common weight policy without having multiple connections. The results are based on a random regular network with 100 nodes, each with degree 3. In the second phase, we use the same network but each IoT device is assumed to have multiple connections to initiate the payment. 

\begin{figure*}[!htb]
\begin{centering}
\begin{tabular}{ccc}
\includegraphics[keepaspectratio=true,angle=0,width=53mm]{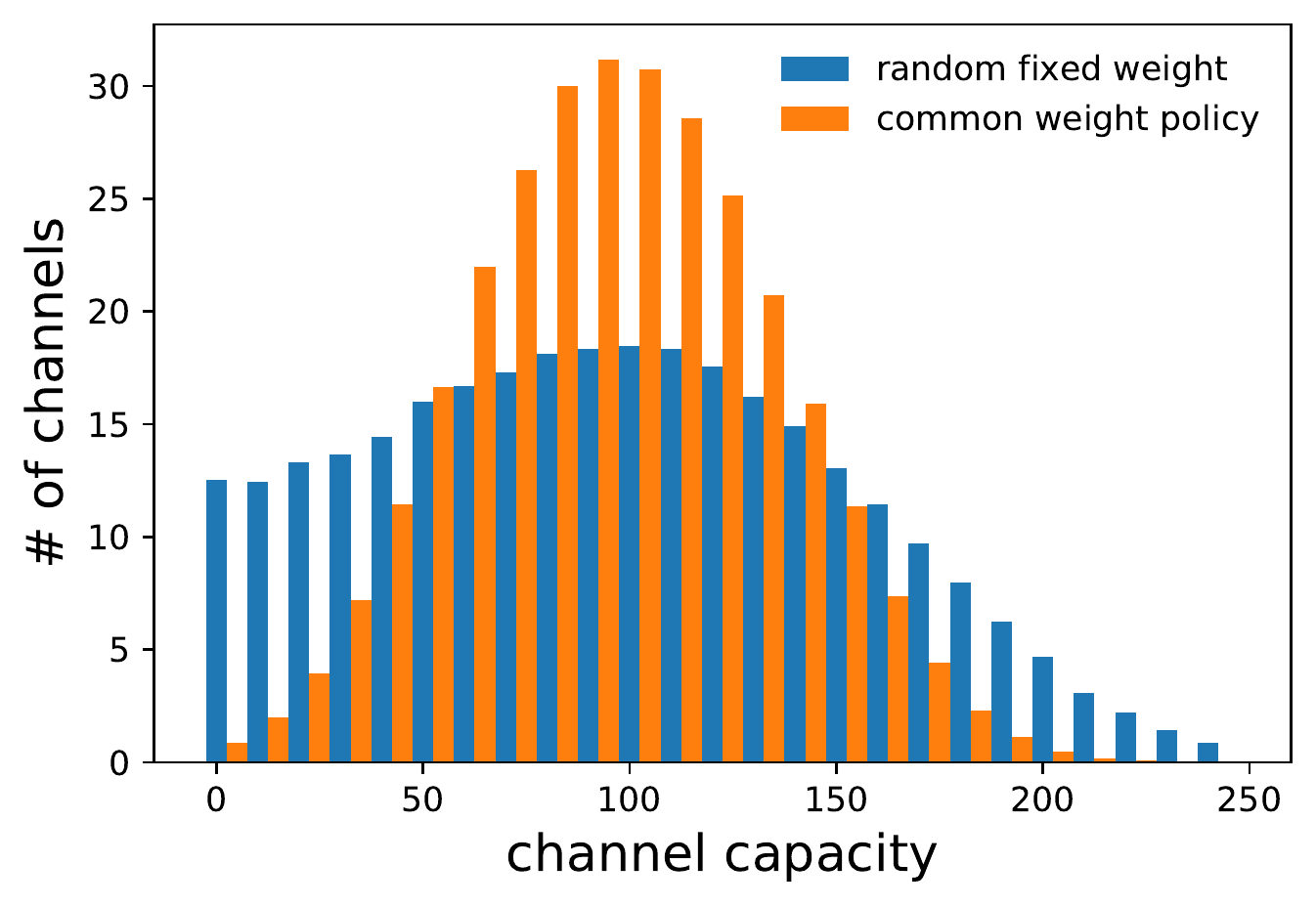} &
\hspace{-5mm}
\includegraphics[keepaspectratio=true,angle=0,width=56mm]{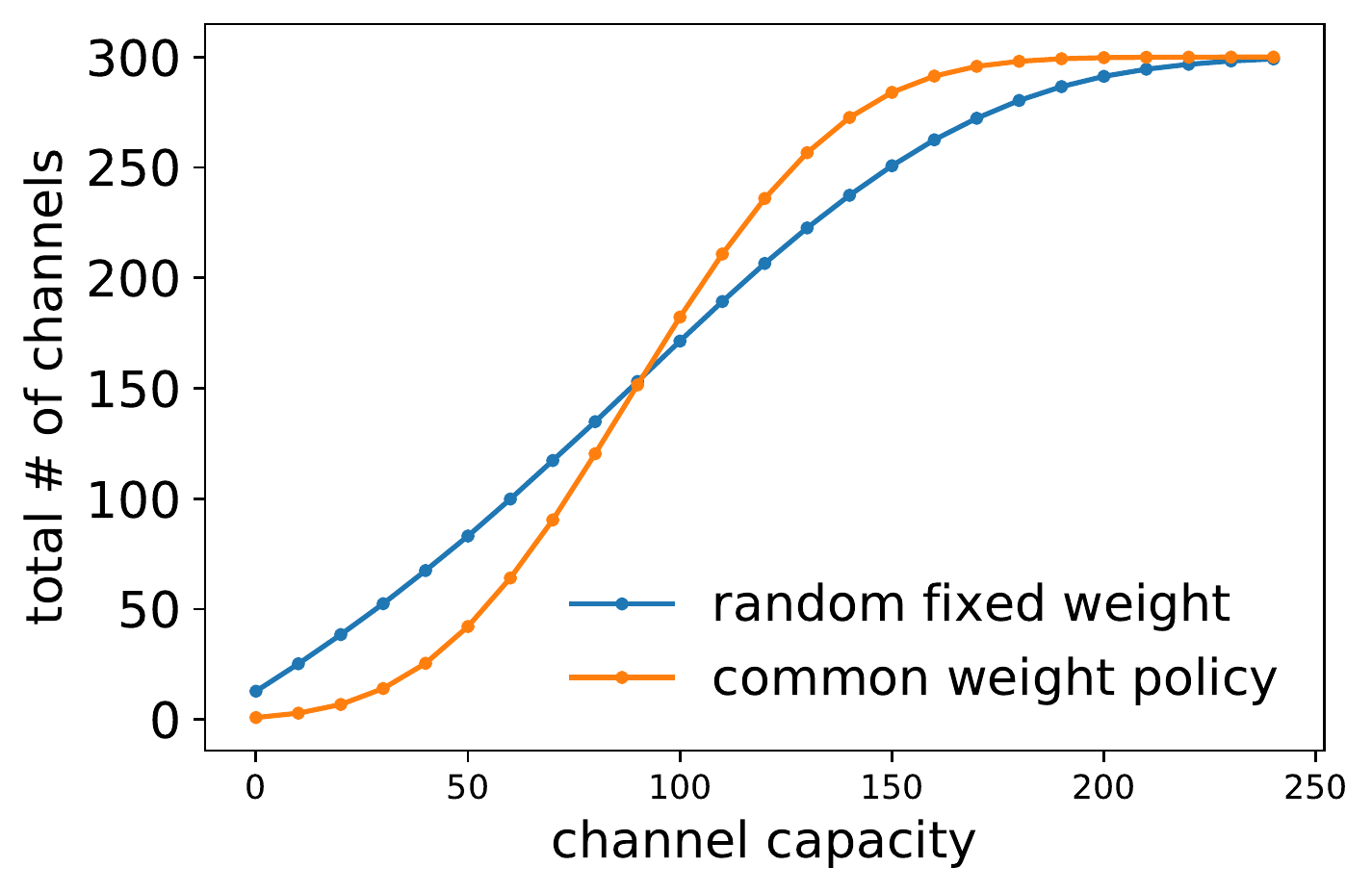} &
\hspace{-5mm}
\includegraphics[keepaspectratio=true,angle=0,width=53mm]{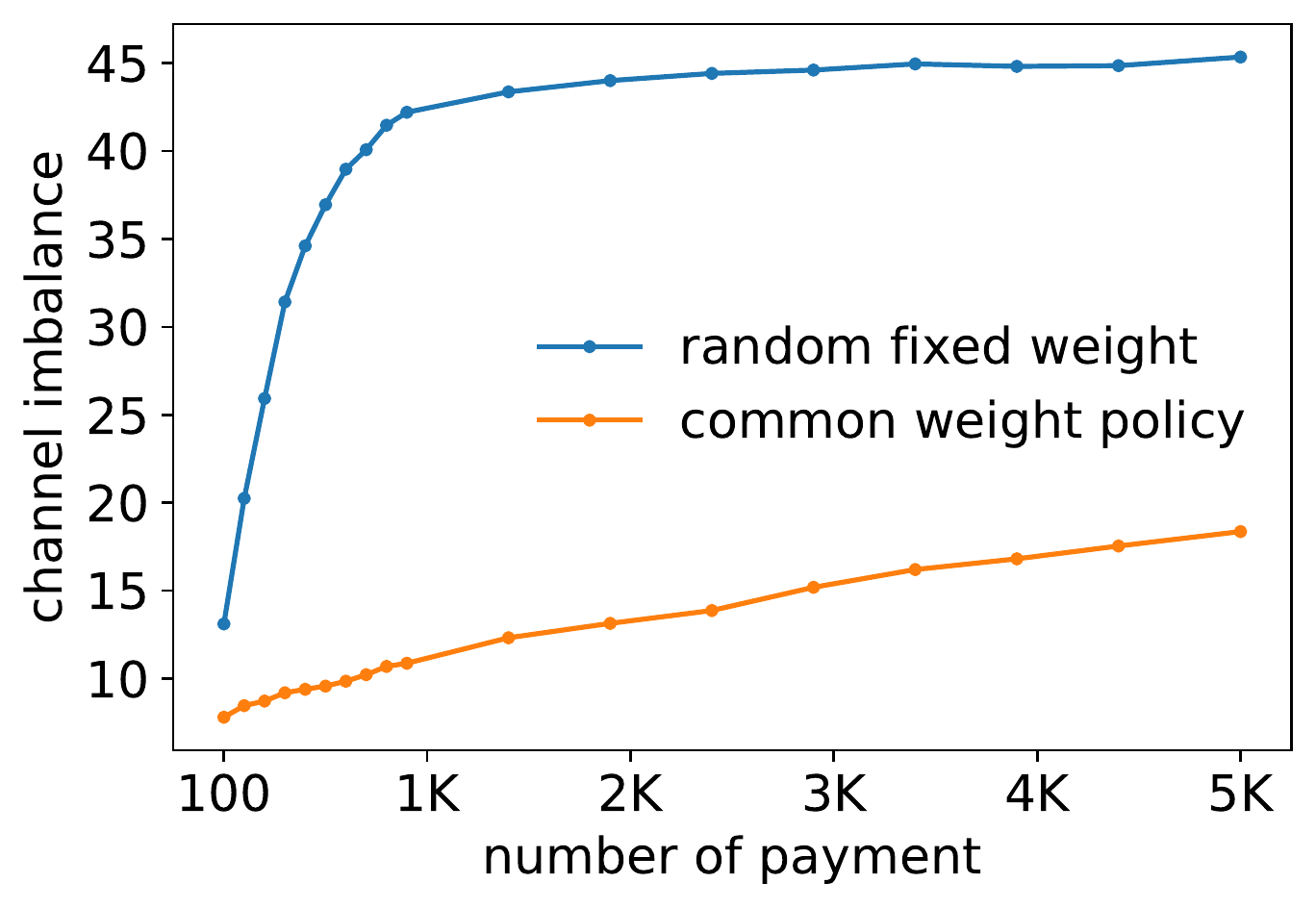} 

\vspace{-2mm}

 \tabularnewline
\footnotesize{ a. Capacity Distribution} &
\hspace{-5mm}
\footnotesize{ b. Capacity Cumulative Distribution } & 
\hspace{-5mm}
\footnotesize{ c. Network Imbalance } 
\vspace{0mm}
 
\tabularnewline

\includegraphics[keepaspectratio=true,angle=0,width=53mm]{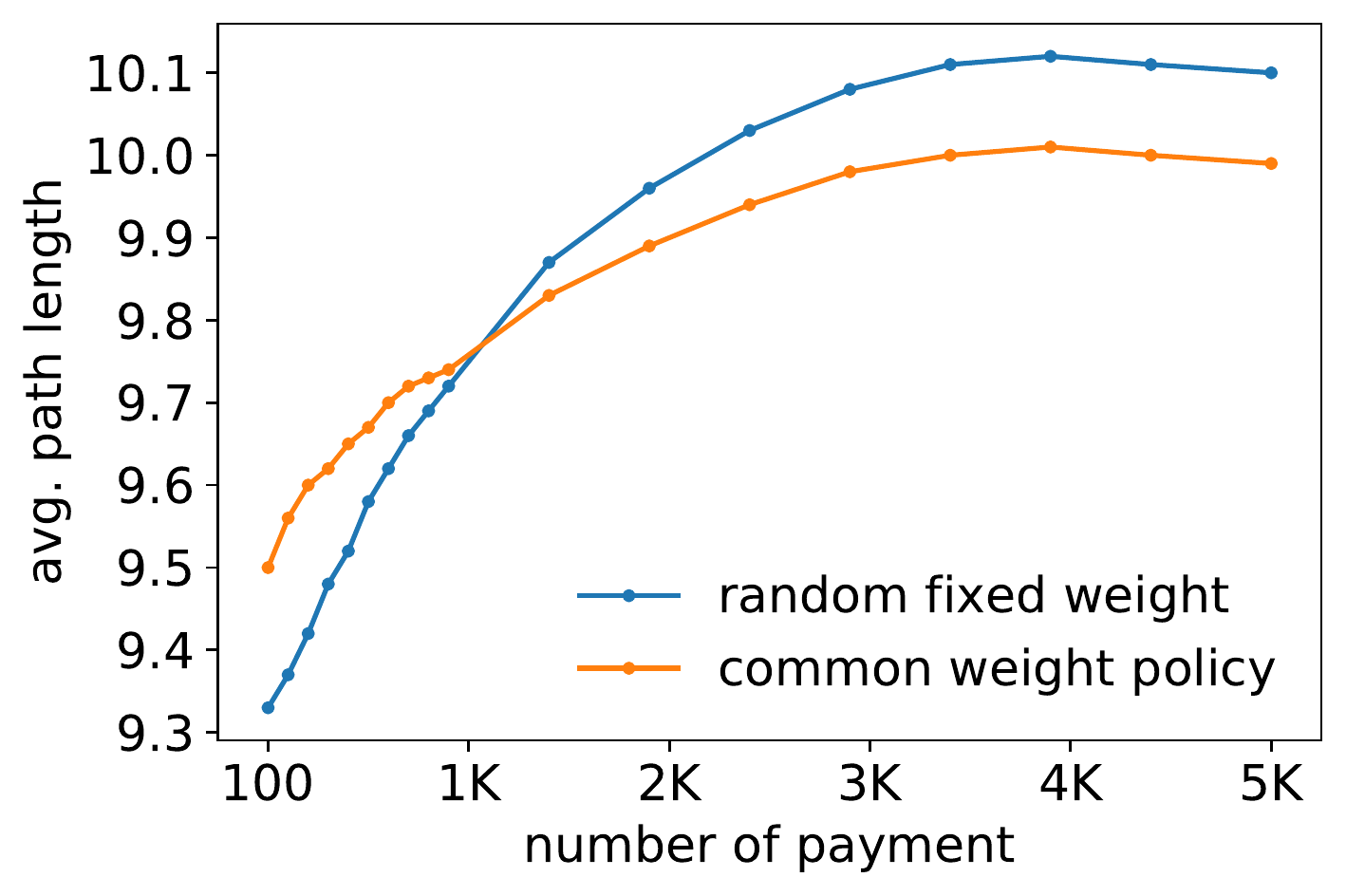} & 
\hspace{-5mm}
\includegraphics[keepaspectratio=true,angle=0,width=52mm]{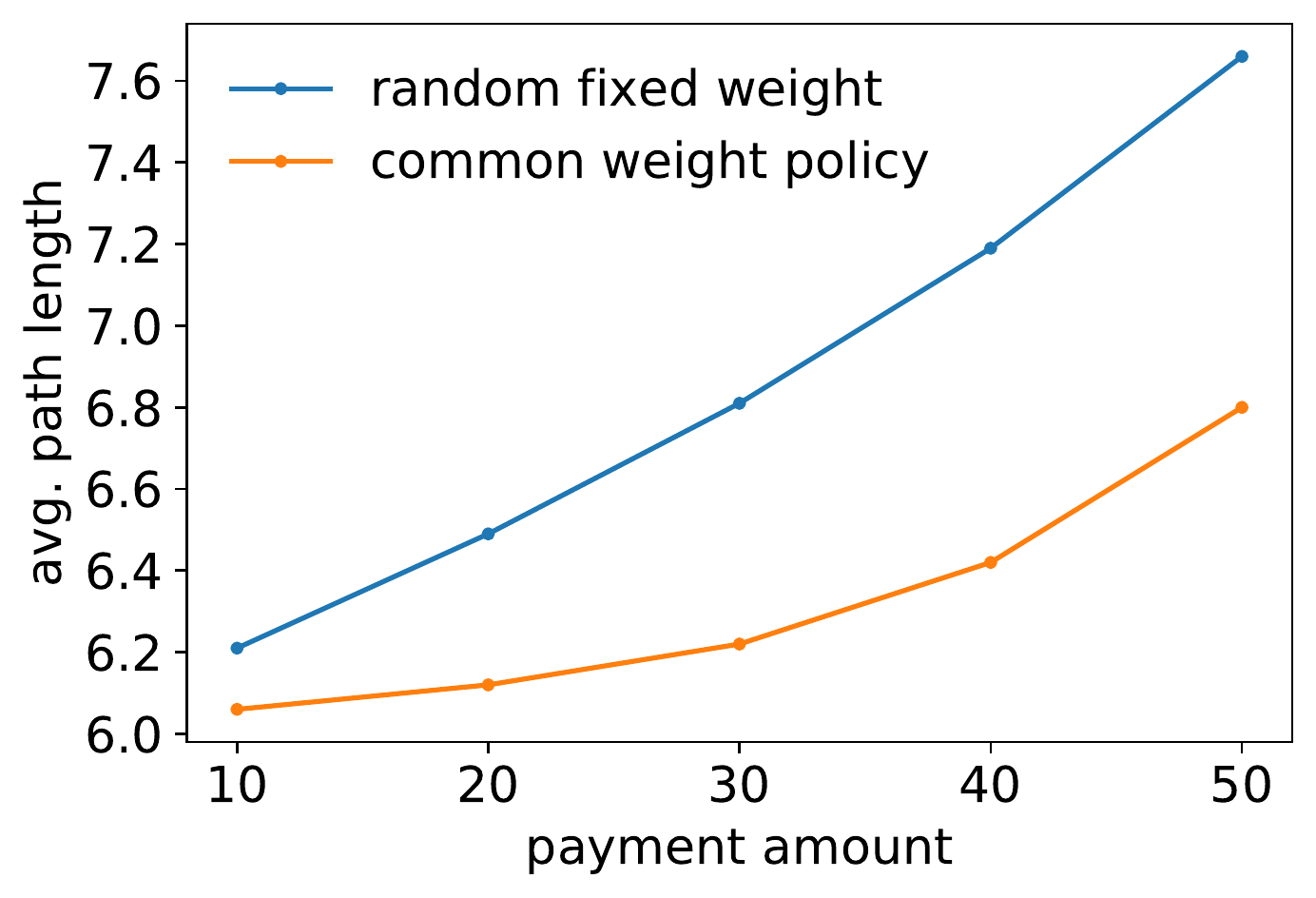} & 
\hspace{-5mm}
\includegraphics[keepaspectratio=true,angle=0,width=52mm]{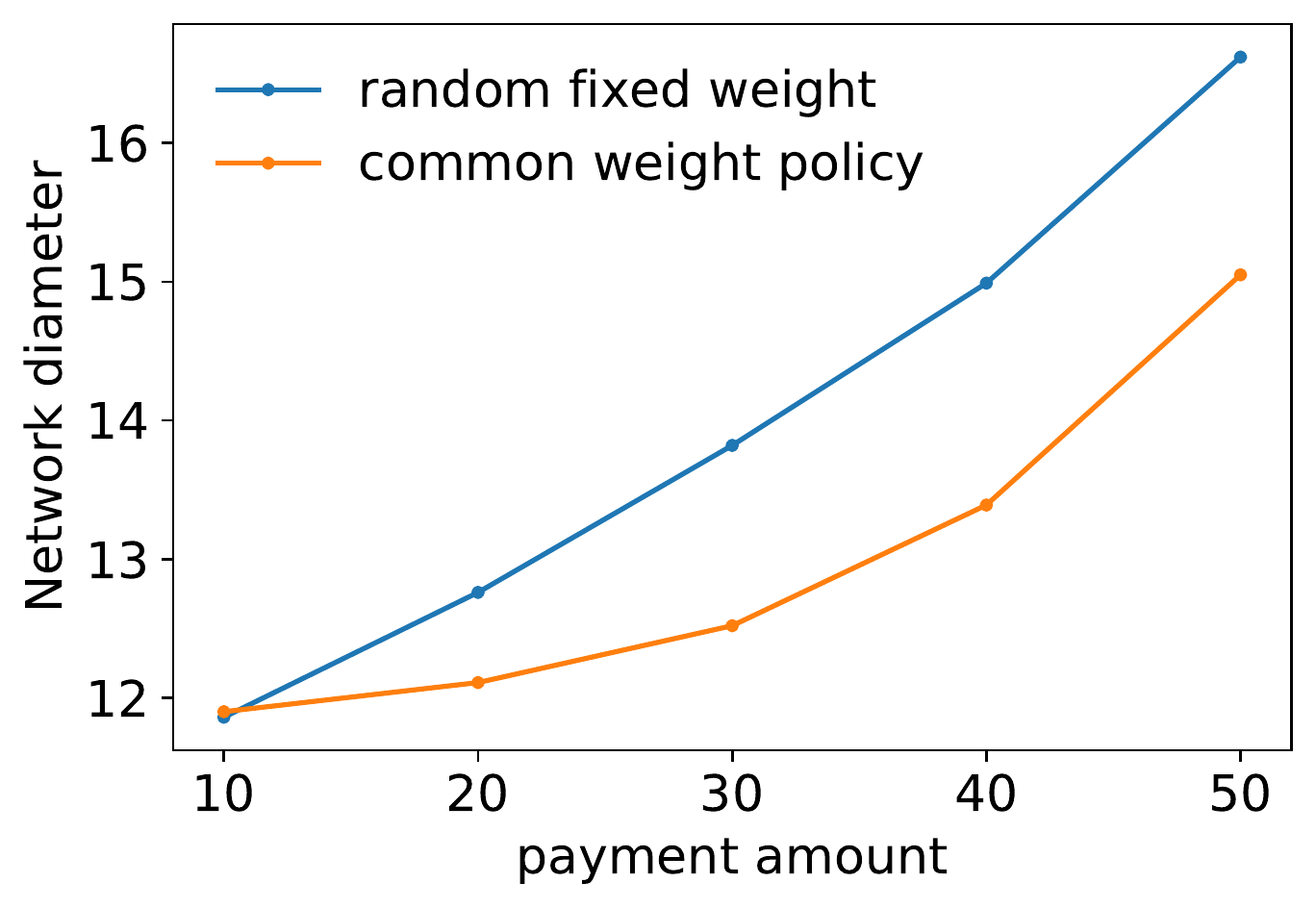} 
\vspace{-2mm}
 \tabularnewline
\footnotesize{ d. Moving average of payment length} & 
\hspace{-5mm}
\footnotesize{ e. Average Path Length } &
\hspace{-5mm}
\footnotesize{ f. Network Diameter }
\vspace{0mm}
 \tabularnewline

\end{tabular}
\caption{Experiment Results of Common Weight using Single Connection}
\label{fig:res1}
\vspace{-6mm}
\end{centering}
\end{figure*}

\textit{Payment files:} For the first set, we built 10 different balanced payment sequence consisting of 5000 end-to-end transactions. The number of payments is equally distributed among nodes. Each node sends and receives 50 transactions between \$5 and \$15. So, total incoming and outgoing may not be the same but expected to be close since they are picked randomly. Source and destination are not necessarily the same which means that node $A$ sends to $B$ but receives from $C$. The payment sequences in the second set are unbalanced which allows us to investigate the impact of multi-connection.

\textit{Payment transfer:} Each node calculates the path and the payment is sent through intermediate nodes by decrementing the amount from each channel used and incrementing in the opposite direction.

\textit{Experiment run:} Each payment file is run with 100 different seeds which randomized the initial channel capacity and payment amount. So, the results are aggregate of 1K randomized tests on the network.

\subsection{Performance Metrics}

We use the following metrics to assess the performance of the proposed approach: 

\begin{itemize}
    
    \item\textit{Network imbalance:} This metric shows the deviation of channel balances from the overall average. It is defined as the average of difference of each channel to average capacity.

    \item\textit{Path length:} This is the number of hops that an actual payment has traveled.


    \item\textit{Network diameter:} This is the minimum hop count between two most distant nodes.

    \item\textit{Success ratio:} This metric shows the number of payments that could be sent successfully from a source to destination for a node.

\end{itemize}

For comparison, we developed a random fixed weight policy where the channel weights among the nodes are fixed to an initial value. The least cost paths are found using these random weights.

\subsection{Experiment Results}

\subsubsection{Single Connection Experiments}
We first present the results that have been collected from the experiment where we applied common weight policy using a single connection from each IoT device to store. 
Fig~\ref{fig:res1}.a and Fig~\ref{fig:res1}.b show the channel capacity distribution.  A bar represents the number of channels whose capacity is between a given $x$ and $x+10$. There are 300 directional channels, all of which were initially assigned a balance between 50 and 150 and it is uniformly distributed. When we apply random fixed weight, the distribution gets flattened. A substantial part of the channels is low-balanced. A quarter of them drops below 50 which was the lowest value for a channel in the initial setup. 25 channels' capacity is less than 20. In the case of our proposed common weight policy, the channel capacity distribution resembles a Gaussian distribution with a mean 100. 85 \% of the channels is still in the initial range. The number of channels whose balance is less than 20 is only 3. This suggests a more balanced network. 

To measure the deviation more precisely and quantify it, we use the network imbalance metric. We set all the channels to 100 so that we can measure the variance accurately. We preferred to use a timeline in the x-axis to see the change. After every hundred payment, network imbalance is measured. As shown in Fig~\ref{fig:res1}.c, for the first case, it increases dramatically, then continues increasing slowly. For our approach, the value is much less and the increase is steady. It ends up with an average distance of 18. It is obvious that the channel capacities are staying closer to mean.

\begin{figure*}[!htb]
\begin{centering}
\begin{tabular}{ccc}

\includegraphics[keepaspectratio=true,angle=0,width=53mm]{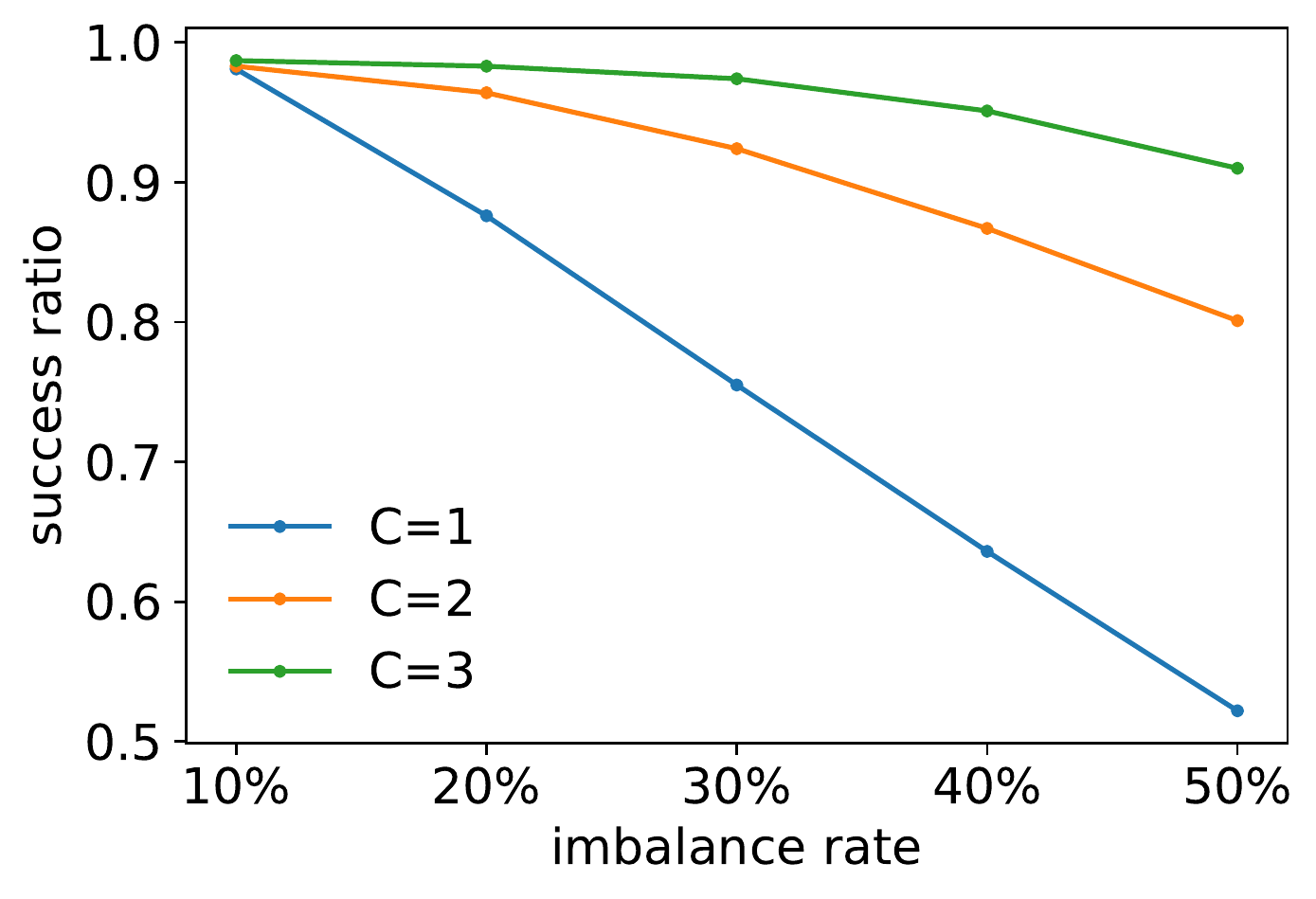} &
\hspace{-5mm}
\includegraphics[keepaspectratio=true,angle=0,width=53mm]{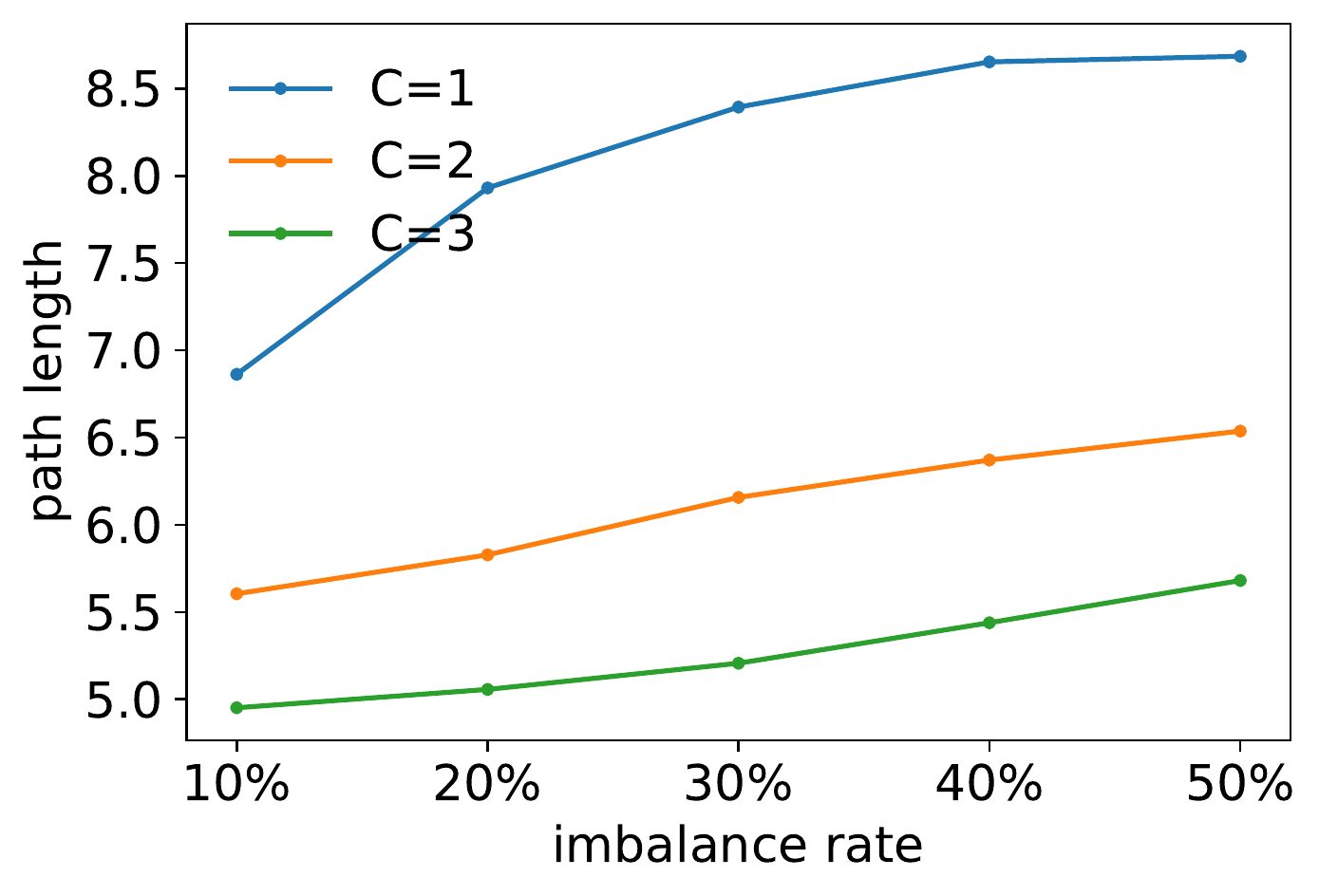} &
\hspace{-5mm}
\includegraphics[keepaspectratio=true,angle=0,width=53mm]{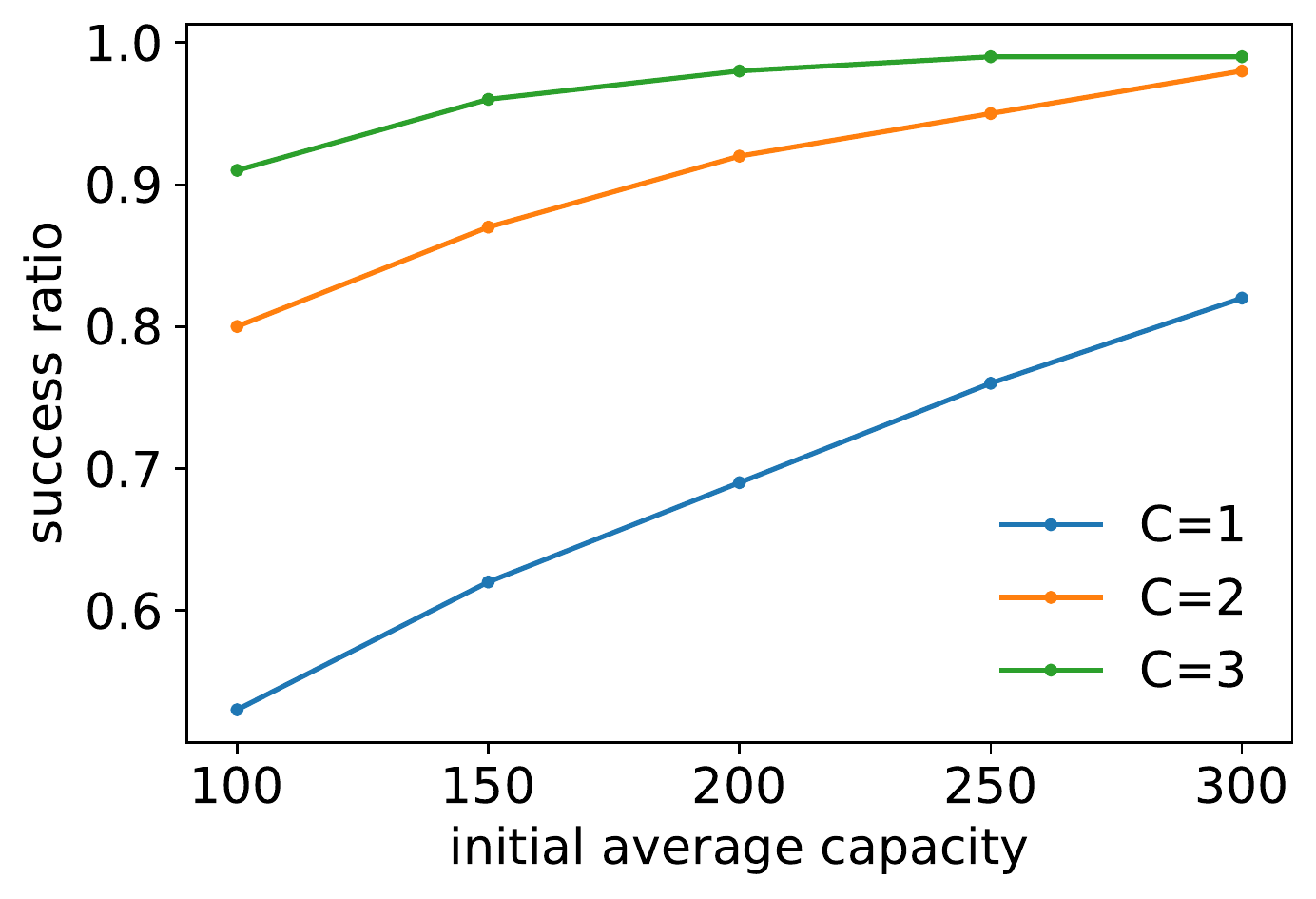} 

\vspace{-2mm}

 \tabularnewline
\footnotesize{ a. Success ratio vs. Imbalance rate} &
\hspace{-5mm}
\footnotesize{ b. Path length vs. Imbalance rate } & 
\hspace{-5mm}
\footnotesize{ c. Success ratio vs. Initial capacity } 

 \tabularnewline
\end{tabular}
\caption{Experiment Results of Multiple Connection}
\label{fig:res3}
\vspace{-5mm}
\end{centering}
\end{figure*}

To assess whether our approach brings any overhead to the payment performance, we did additional tests. We investigated the impact on path length for payments and network diameter on the network topology. First, we checked the number of hops each payment in the payment file has to go through. For this experiment, we used a network with 1000 nodes to magnify the results. As seen in  Fig~\ref{fig:res1}.d, the quantitative difference between the results of the two methods is not significant. The trends are also similar. It increases with time, then becomes steady. This can be explained as follows: At the beginning, our approach uses longer paths because we force payments to travel over high capacity channels even though they are longer. Fixed weight approach takes advantage of shorter paths at the beginning at the cost of balance exhaustion. This causes payments to go over longer paths at later stages. 

 Indeed, later when we check the average path length among all the nodes in the network after all payments are sent to their destinations, the above observation we made was confirmed. Fig~\ref{fig:res1}.e shows the average path length from each node to every other node in the network for various amounts. Note that the path between two nodes might be different based on the amount. Higher amounts have to travel longer paths. Our approach provides better connectivity over the network which can transfer \$50 from one point to another using one less hop on average. Fig.~\ref{fig:res1}.f shows the diameter of the network for various payment amounts. Applying common weight creates a more compact network. So overall, there is no overhead but actually gain in the long run.

\subsubsection{Multiple Connection Experiments}
In the first batch of the experiments, the payments among the nodes are well distributed which means that the number of payments a node sends and receives is same even though the amount is chosen randomly in a specific range. This enabled us to observe the impact of weight policy without getting into success rate discussions. In this subsection, we perform additional experiments for the cases where IoT devices have multiple connection points to stores. We generated scenarios where the payments sent and received for a particular node is not equally distributed. Specifically, when the payments are skewed, the number of incoming and outgoing payments will not be equal for a node, and thus unsuccessful payments occur because of channel depletion. 

To create imbalance among payments sent and received, we varied the difference among these from 10\% to 50\%. For instance, if the difference is 10\%, then the number of payments sent will be 10\% more or less than the number of payments received. 
We tested single (C1), double (C2) and triple (C3) connections against varying imbalance rates (10 to 50). 

As seen in Fig.~\ref{fig:res3}.a, the success ratio for single connection (C1) drops to 50\% while it is around 90\% for triple connection (C3). Additional connections enable an IoT device to use an alternative route in case one store is disconnected because of outgoing capacity erosion, and re-balance asymmetric channels.  Fig.~\ref{fig:res3}.b shows the average path length of payments for each connection case. These results indicate that multi-connections also reduce the total number of hops that a payment has to go through. The last figure, Fig.~\ref{fig:res3}.c, shows the relationship between the success ratio and the initial capacity of channels when they were established. We varied the average capacity from 100 to 300 for the channels, and the success ratio gets closer to 1 especially with multiple connections.

\section{Conclusion and Future Work}

Cyrptocurrency payment networks are newly emerging and promising area which requires more investigation to make them more efficient and effective. In this paper, we proposed and investigated the adoption of a network-wide common weight policy and redundant connections between IoT devices and stores for cryptocurrency payment networks. The proposed approach aims to keep the links balanced in payment channel network by diverting the payments towards high-capacity connections. The results show that we are able to sustain the network more equally balanced and obtain higher success ratio under skewed payment flows. 
\vspace{-2mm}
\bibliographystyle{IEEEtran}


\end{document}